\documentclass[aps,prl, reprint,superscriptaddress]{revtex4-2}
\usepackage{graphicx}
\usepackage{dcolumn}
\usepackage{bm}

\usepackage[utf8]{inputenc} 
\usepackage[T1]{fontenc}    
\usepackage{hyperref}       
\usepackage{url}            
\usepackage{booktabs}       
\usepackage{amsfonts, amsmath, amssymb, amsopn}       
\usepackage{nicefrac}       
\usepackage{microtype}      
\usepackage{lipsum}
\usepackage{graphicx}       
\usepackage{comment}
\begin{document}

\title{Mutual Information in High-Dimensional Neural Coding}

\author{Zimo Li}
\affiliation{Interdepartmental Neuroscience Program, Yale University, New Haven, CT, USA}
\affiliation{Quantitative Biology Institute, Yale University, New Haven, CT, USA}
\affiliation{Department of Neuroscience, Yale University, New Haven, CT, USA}

\author{Yuchang Tian}
\affiliation{Department of Statistics and Data Science, Yale University, New Haven, CT, USA}

\author{Matthew P. Leighton}
\altaffiliation{These authors jointly supervised this work. Contact author: matthew.leighton@yale.edu, christopher.lynn@yale.edu}
\affiliation{Quantitative Biology Institute, Yale University, New Haven, CT, USA}
\affiliation{Department of Physics, Yale University, New Haven, CT, USA}

\author{Christopher W. Lynn}
\altaffiliation{These authors jointly supervised this work. Contact author: matthew.leighton@yale.edu, christopher.lynn@yale.edu}
\affiliation{Quantitative Biology Institute, Yale University, New Haven, CT, USA}
\affiliation{Department of Physics, Yale University, New Haven, CT, USA}
\affiliation{Wu Tsai Institute, Yale University, New Haven, CT, USA}

\date{\today}

\begin{abstract}
Estimating mutual information in large neural populations is limited by the exponential growth of possible neural response patterns. 
Here we apply a Monte Carlo framework to make this problem tractable 
by sampling from a conditionally independent population model fit from single-neuron responses. The estimator accurately recovers analytical ground truth, corrects finite-data bias, and scales to empirical recordings containing $10^2$ to $10^4$ neurons from visual cortex, hippocampus, and retina. Across systems, stimulus uncertainty decays exponentially with population size, revealing a simple scaling law for high-dimensional neural codes.
\end{abstract}

\maketitle

\begin{figure*}
\includegraphics{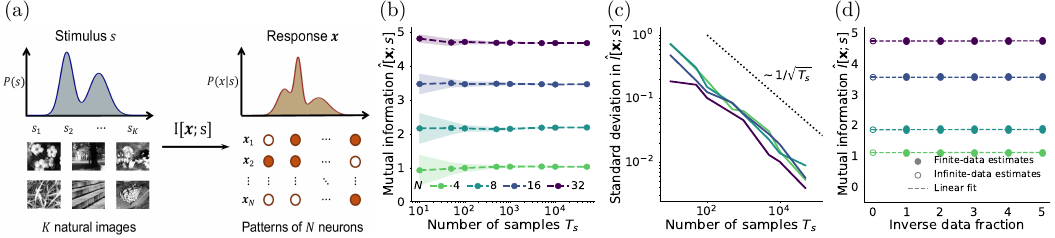}
\caption{\label{fig:fig1} Monte Carlo sampling for mutual information estimation.
(a) Illustration of neural coding, where $K$ stimuli (here natural images) are repeatedly presented, and neural responses are recorded from $N$ neurons.
(b) Convergence of mutual information as a function of the number of Monte Carlo samples $T_s$ for artificial neurons responding to $K = 32$ stimuli with firing probabilities drawn uniformly from $[0,1]$. Lines and shaded regions represent means and standard deviations over ten sets of Monte Carlo samples.
(c) Fluctuations in mutual information estimates decrease as $1/\sqrt{T_s}$.
(d) Mutual information estimates as a function of the inverse fraction $1/f$ of the experimental data. Dashed lines represent linear fits, and infinite-data estimates are computed by extrapolating to $1/f = 0$.
}
\end{figure*}

A central goal of neuroscience is to understand how neural activity encodes external variables ~\cite{Bialek1991, Calderbank1998, Panzeri1999, Borst1999, Knill2004, Sharpee2004, Naselaris2011, Barlow2012, Craik2012}.
The neural response to a stimulus can be formulated probabilistically: 
each stimulus $s$ produces a neural response $\bm{x} = \{x_i\}$ drawn from a conditional distribution $P(\bm{x}|s)$, where $x_i$ is the response of neuron $i$. Mutual information $I[\bm{x};s]$ measures the strength of the relationship between stimulus and response: it quantifies how much the neural activity reduces our uncertainty about the stimulus
[Fig.~\ref{fig:fig1}(a)]~\cite{Schneidman2003, Mathis2012,Barry2013, Pachitariu2015, Huang2018, Molter2020,Pohl2026}.
Consequently, it provides the foundation for theories of neural coding, 
including coding efficiency~\cite{Fairhall2001, Laughlin2001, Lewicki2002, Smith2006, Adibi2013, Liu2026}, maximizing entropy~\cite{Jaynes1957,Carcamo2025, Lynn2025}, 
feature representation~\cite{Simoncelli2001, Louie2012, Mathis2012}, 
and the information bottleneck~\cite{Palmer2015, Sachdeva2021}.

Modern experiments now provide measurements of the activity of thousands of neurons simultaneously~\cite{Stosiek2003, Cheng2011, Jun2017, Steinmetz2021, Paulk2022, Zong2022}. However, as recordings increase in size, the number of possible responses explodes exponentially, posing two challenges in the estimation of mutual information.
First, one must define or estimate the distribution of neural responses for each stimulus $P(\bm{x}|s)$.
Even in the simplest scenario where each neuron is binary ($x_i = 1$ for active or $x_i = 0$ for silent), 
there are still $K2^N$ different stimulus-response pairs, 
where $N$ is the size of the population and $K$ is the number of stimuli. 
As a result, directly estimating the response probabilities $P(\bm{x}|s)$ is only possible in small populations of $N\sim 10$ cells.
Second, one must compute the expectation of $\log [P(\bm{x}|s)/P(\bm{x})]$ over all stimuli and responses.
While both challenges stem from the exponential growth of possible responses, 
we will see that only the former fundamentally limits our ability to reliably estimate mutual information.

Existing methods for estimating mutual information in neural populations can be organized by how they confront the curse of dimensionality. 
Direct estimators compute mutual information from observed response counts, often with sparse-pattern approximations such as the singleton estimator~\cite{Kraskov2004, Barry2013, Nardin2023}. 
These methods make few assumptions about the structure of the population code, but are restricted to small populations or to regimes in which the relevant response patterns are sufficiently sampled. 
A second approach is to introduce explicit structure into the population distribution using maximum-entropy models, latent-variable models, or other parametric stimulus-response models~\cite{Christensen1985, GranotAtedgi2013, Tkacik2014, Carcamo2025}. 
Such models can reveal how correlations affect coding, but computing $P(\bm{x}|s)$ still requires normalization, which is typically only possible for small networks. 
A third strategy avoids the full response space by estimating information after dimensionality reduction, or by training neural-network estimators to approximate information-theoretic objectives~\cite{KochJanusz2018, Gowri2024, Abdelaleem2025}. 
These approaches are powerful in settings where the relevant dependence is low-dimensional or where large training sets are available.

Here, we study high-dimensional mutual information in large neural populations under the assumption of conditional independence. 
As the natural first-order model for $P(\bm{x}| s)$, conditional independence
reduces the intractable challenge of estimating probabilities for all $K2^N$ stimulus-response pairs to the manageable problem of estimating $N\times K$ single-neuron firing probabilities $P(x_i| s)$. 
The remaining obstacle in computing the expectation of \(\log P(\bm{x}| s)/P(\bm{x})\) can be overcome with a Monte Carlo estimator that samples stimulus-response pairs from the conditionally independent model and averages the information density directly~\cite{Yarrow2012,Huang2018,reinhardt2023path}. 
We validate this approach using an analytically tractable model for which the ground-truth mutual information can be computed exactly, establishing reliable convergence with only a modest number of samples. 
Applying this method to large-scale recordings from mouse visual cortex, mouse hippocampus, and salamander retina, 
we find that stimulus uncertainty decays exponentially with the number of neurons. These results confirm existing theoretical predictions \cite{kang2001mutual} and provide a starting point for estimating mutual information in neural populations far beyond the reach of direct enumeration. \\

\begin{figure*}
\includegraphics{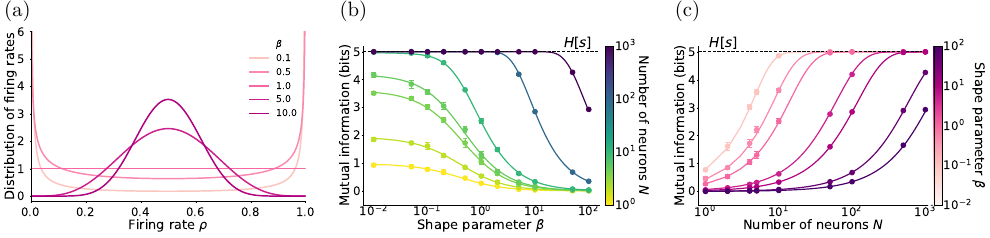}
\caption{\label{fig:fig2} Monte Carlo estimates recover analytic mutual information.
(a) Symmetric beta distributions over firing rates for different shape parameters $\beta$.
(b) Mutual information as a function of $\beta$ for different population sizes $N$. 
(c) Mutual information as a function of the number of neurons $N$ for different shape parameters $\beta$. In (b) and (c), lines are analytic predictions from Eq.~(\ref{eq:Iana}), and points with error bars represent mean $\pm$ standard deviation across ten Monte Carlo estimates. Dashed lines indicate the stimulus entropy $H[s] = 5$~bits.
}
\end{figure*}

\emph{Estimating mutual information}---We aim to estimate the mutual information between high-dimensional neural activity $\bm{x}$ and a discrete stimulus $s$,
\begin{equation}
\label{eq:I}
I[\bm{x};s] = \sum_s\sum_{\bm{x}}P(\bm{x}, s)\log \frac{P(\bm{x}|s)}{P(\bm{x})}.
\end{equation}
If the number of stimuli is relatively small, or if stimuli are imposed experimentally, then we have access to the stimulus distribution $P(s)$. For large populations, however, we do not have access to the response distribution $P(\bm{x}|s)$. Instead, we can measure the firing probability $P(x_i = 1| s)$ of each neuron $i$ in response to each stimulus $s$. With this knowledge and nothing else, the collective response of the population is defined by the conditionally independent model,
\begin{equation}
\label{eq:PCI}
P(\bm{x}|s)=\prod_{i=1}^{N}P(x_i|s).
\end{equation}

To compute the mutual information, we cannot sum over all responses and stimuli in Eq.~(\ref{eq:I}). Instead, we iteratively sample stimuli $s$ from the marginal $P(s)$, and for each stimulus, sample a corresponding response $\bm{x}$ from the model $P(\bm{x}|s)$. For each sampled response, we evaluate $P(\bm{x}|s)$ and the marginal probability $P(\bm{x})=\sum_s P(s)P(\bm{x}|s)$ to compute the log-ratio $\log\left[\frac{P(\bm{x}|s)}{P(\bm{x})}\right]$. Repeating this process for $T_s$ stimulus-response samples, we arrive at a Monte Carlo estimate for the mutual information ~\cite{Yarrow2012, Huang2018},
\begin{equation}
\hat{I}_{T_s} = \frac{1}{T_s}\sum_{t=1}^{T_s}\log\left(\frac{P(\bm{x}(t)|s(t))}{P(\bm{x}(t))}\right). 
\end{equation}

To study large populations, we need the estimate $\hat{I}_{T_s}$ to converge for a number of samples $T_s$ much smaller than the number of terms in Eq.~(\ref{eq:I}). Consider artificial neurons with firing probabilities $P(x_i = 1|s)$ drawn from the uniform distribution on $[0,1]$. As the number of samples $T_s$ increases, we find that the estimate $\hat{I}_{T_s}$ quickly converges to a stable value, with fluctuations decaying as $1/\sqrt{T_s}$ [Fig.~\ref{fig:fig1}(b)-(c)]. We confirm the same convergence in multiple large-scale recordings (Supplemental Material). Thus, the limiting step in estimating high-dimensional mutual information is not summing over responses and stimuli; it is having access to normalized probabilities $P(\bm{x}|s)$.

In previous work, this Monte Carlo estimator has been applied to models of artificial neurons~\cite{Yarrow2012, Huang2018}. In what follows, we investigate the mutual information of real neurons under conditional independence. For a given population, the model in Eq.~(\ref{eq:PCI}) is not defined with infinite precision, but instead must be estimated from $T_f$ samples of stimulus responses measured in experiments. In the Supplemental Material, we show that finite data leads to a bias in the mutual information estimate that scales as $K(N+1)/(2T_f)$.

To account for this bias, following existing finite-data correction techniques~\cite{Treves1995, Strong1998, Steuer2002, Palmer2015},
we subsample different fractions $f$ of the data. For each subsample, we 
estimate the empirical firing probabilities $P(x_i|s_j)$
and then repeat the mutual information estimation using  $T_s=10^4$ Monte Carlo samples.
Plotting these finite-data estimates against the inverse data fraction $1/f$, we can use linear regression to extrapolate to infinite data at $1/f = 0$. Repeating this entire process, we arrive at multiple infinite-data estimates, from which we compute the average and standard deviation of the corrected mutual information [Fig.~\ref{fig:fig1}(d)]. In all settings, finite-data corrections are small, which gives us confidence in our mutual information estimates (Supplemental Material). \\

\emph{Minimal model of neural coding}---To assess the accuracy of the Monte Carlo estimator for high-dimensional mutual information, 
we study an analytically tractable model of neural coding.
For each neuron $i$ and stimulus $s$, the firing probability $P(x_i=1|s)$ is drawn independently from a symmetric beta distribution $\mathrm{Beta}(\beta,\beta)$. 
When $\beta=1$, this reduces to the uniform distribution from Fig.~\ref{fig:fig1}.
For $\beta \ll 1$, neuronal firing becomes deterministic; and when $\beta \gg 1$, firing probabilities concentrate around $1/2$, such that neurons decouple from the stimuli [Fig.~\ref{fig:fig2}(a)]. We therefore expect mutual information to decrease as $\beta$ increases.

This model permits an exact solution for the mutual information,
\begin{equation}
\label{eq:Iana}
I[\bm{x};s] =\log K - 2^N\int_0^{\infty}\frac{dt}{t}\phi'_N(t)[\phi_N(t)^{K-1}-1],
\end{equation}
where
\begin{equation}
\phi_N(t) = \sum_{k=0}^{\infty}\frac{(-t)^k}{k!}\left<\rho^k\right>^N,
\end{equation}
and $\langle \rho^k\rangle$ is the $k^\text{th}$ moment of the beta distribution over the firing probabilities $\rho$ (Supplemental Material). For different values of $\beta$, we consider different numbers of neurons $N$ responding to $K=32$ stimuli. For each population, we simulate $T_f=10^5$ stimulus presentations and population responses. From these, we infer the firing probabilities $P(x_i=1|s_j)$ and estimate mutual information by Monte Carlo sampling. Across all population sizes, mutual information decreases with $\beta$ [Fig.~\ref{fig:fig2}(b)], and across all shape parameters, mutual information increases with the number of neurons $N$ [Fig.~\ref{fig:fig2}(c)]. In all cases, the Monte Carlo estimates closely track the analytic mutual information, even for large $N$ and strong coding $\beta \ll 1$. \\

\begin{figure*}
\includegraphics{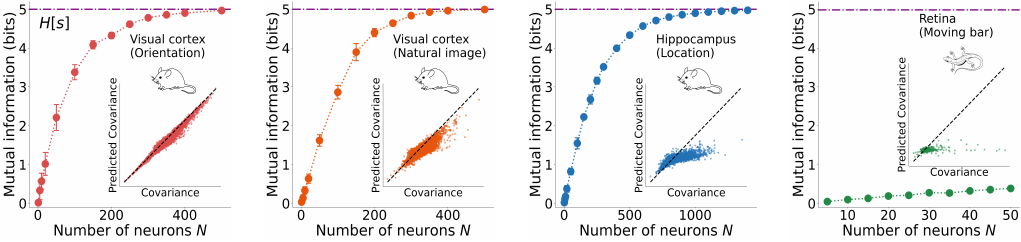}
\caption{\label{fig:fig3} Mutual information in large populations.
(a) Estimated mutual information as a function of the number of neurons $N$ subsampled from mouse primary visual cortex under stimulation with orientation gratings.
(b) Mutual information as a function of the number of neurons $N$ subsampled from mouse primary visual cortex under stimulation with natural images.
(c) Mutual information as a function of the number of neurons $N$ subsampled from mouse hippocampus as the animal runs around a circular track.
(d) Mutual information as a function of the number of neurons $N$ subsampled from the retina of a salamander responding to an artificial moving bar stimulus.
In all panels, mutual information is estimated using the Monte Carlo estimator under conditional independence. Points and error bars denote mean \(\pm\) standard deviation across ten randomly selected neuron groups. Dashed lines indicate the stimulus entropy $H[s]$, which is set to 5 bits for $K = 32$ equally-probable stimuli.
The number of neurons is increased until the mutual information increases to within $1\%$ of the stimulus entropy.
Insets illustrate the covariances between neurons measured in data versus those predicted by the conditional indepedent model; dashed lines indicate equality.}
\end{figure*}

\emph{High-dimensional neural coding}---We are now prepared to study mutual information in large populations of neurons. We first consider $N=8,991$ neurons in the mouse visual cortex, recorded using calcium imaging in previous experiments as the animal observes $K=32$ different orientation gratings~\cite{Stringer2019}. After preprocessing, each neuron is classified as active ($x_i = 1$) or silent ($x_i = 0$) for each stimulus presentation (Appendix). In this population, we find that the conditionally independent model [Eq.~(\ref{eq:PCI})] quantitatively predicts the correlations between neurons [Fig.~\ref{fig:fig3}(a), \textit{inset}], suggesting that it provides an accurate description of the population responses.

To understand how mutual information scales with population size, we randomly subsample neurons and, for each subpopulation, compute the finite-data-corrected mutual information. Just as for synthetic neurons, we find that the Monte Carlo estimator quickly converges with the number of samples $T_s$ and that finite data corrections are small (Supplemental Material). The mutual information increases rapidly with the number of neurons $N$ [Fig.~\ref{fig:fig3}(a)]. Only 63 neurons are needed to halve the uncertainty about the stimulus, quantified by the conditional entropy $H[s|\bm{x}] = H[s] - I[\bm{x};s]$. With 613 neurons, the mutual information reaches within 1\% of its maximum possible value, the stimulus entropy $H[s]$.

We next investigate $N= 11,523$ neurons in the mouse visual cortex, recorded in the same experiment as the animal observes $K =32$ natural images~\cite{Stringer2019}. Compared to the orientation gratings, the conditionally independent model captures less of the correlation structure in the population [Fig.~\ref{fig:fig3}(b), \textit{inset}]. However, we find that the mutual information grows at a similarly rapid rate with the number of neurons $N$ [Fig.~\ref{fig:fig3}(b)]. With $N=71$ neurons, uncertainty about the stimulus is reduced by half, and 513 neurons are sufficient to raise the mutual information to within 1\% of the stimulus entropy [Fig.~\ref{fig:fig3}(b)].

While neurons in the visual cortex respond to visual stimuli, in the hippocampus, some neurons are tuned to fire when the animal is in a specific location~\cite{o1978hippocampus, o1971hippocampus}. Consider $N = 1,485$ neurons in the CA1 region of the mouse hippocampus, recorded using calcium imaging as the animal runs around a circular track~\cite{Gauthier2018}. To enable direct comparison with the visual cortex, we discretize the position of the mouse into $K=32$ bins of equal probability. In this context, the conditionally independent model [Eq.~(\ref{eq:PCI})] is defined by the firing rate $P(x_i=1|s)$ of each neuron $i$ as a function of position $s$, known as the place field of the cell. This model captures some, but not all, of the correlations between neurons [Fig.~\ref{fig:fig3}(c), \textit{inset}]. Compared to the visual cortex, the mutual information grows more slowly with population size $N$ [Fig.~\ref{fig:fig3}(c)]. On average, we need $174$ neurons to reduce stimulus uncertainty by half, and even with all the neurons, the mutual information does not reach 1\% of the stimulus entropy. Additionally, we find that varying the resolution with which we discretize space does not qualitatively change the estimated mutual information (Supplemental Material).

Finally, we estimate the mutual information for $N = 53$ neurons in the salamander retina, recorded using a multi-electrode array as the cells respond to an artificial moving bar~\cite{Palmer2015}. To directly compare with the previous populations, we coarse-grain the moving bar stimulus into $K = 32$ bins of equal probability based its position in the visual field. Perhaps surprisingly, the conditionally independent model does not capture the strong correlations between neurons [Fig.~\ref{fig:fig3}(d), \textit{inset}]. Consequently, the information encoded by retinal neurons is lower than that observed in the mouse cortex and hippocampus [Fig.~\ref{fig:fig3}(d)]. Together, these results demonstrate how, assuming conditional independence, mutual information can be estimated robustly in large populations across distinct brain regions, recording modalities, and stimulus ensembles. \\

\begin{figure*}
\includegraphics{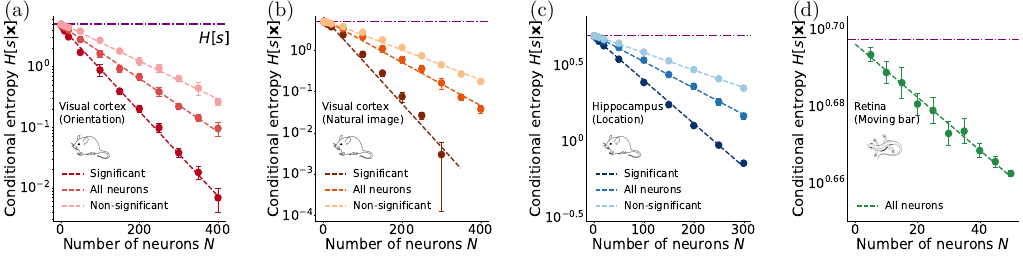}
\caption{\label{fig:fig4} Exponential decay of conditional entropy. 
(a) Conditional entropy $H[s|\bm{x}]$ as a function of the number of neurons $N$ in mouse primary visual cortex under orientation stimulation 
(all neurons: $\lambda = 0.010$, significant neurons: $\lambda=0.016$, non-significant neurons: $\lambda = 0.007$).
(b) Conditional entropy versus $N$ in mouse visual cortex under natural image stimulation (all neurons: $\lambda = 0.012$, significant neurons: $\lambda=0.024$, non-significant neurons: $\lambda = 0.008$).
(c) Conditional entropy versus $N$ in mouse hippocampus as animal runs along a track (all neurons: $\lambda = 0.004$, significant neurons: $\lambda=0.007$, non-significant neurons: $\lambda = 0.003$).
(d) Conditional entropy versus $N$ in salamander retina under artificial bar stimulation ($\lambda = 0.002$).
In all panels, points and error bars denote mean $\pm$ standard deviation across ten randomly selected neuron groups. Lines represent exponential fits.}
\end{figure*}

\emph{Exponential decay of conditional entropy}---
A clear feature shared across all four populations, as well as simulated neurons, 
is the rapid saturation of mutual information with population size. Indeed, previous theoretical results predict that the conditional entropy $H[s|\bm{x}]$ decays exponentially for conditionally independent neurons~\cite{kang2001mutual}. Moreover, the exponent of this decay is the Chernoff distance between response probabilities induced by different stimuli~\cite{kang2001mutual}. Plotting the estimated conditional entropy as a function of $N$, we find a clear exponential decay in all populations (Fig.~\ref{fig:fig4}).

A larger decay exponent $\lambda$ means that fewer neurons are required for the same decrease in stimulus uncertainty $H[s|\bm{x}]$, or, equivalently, for the same increase in mutual information $I[\bm{x};s]$. Intuitively, neurons whose responses vary significantly between stimuli should carry more information. By contrast, neurons with low firing rates or whose responses vary only weakly between stimuli should contribute less to stimulus decoding. In each population, we can divide the neurons into those that respond significantly to a subset of stimuli and those that do not (Supplemental Material). For each group, we sample populations of increasing size $N$ and fit the exponential scaling of the conditional entropy $H[s|\bm{x}]$. Consistent with the above intuition, neurons with significant responses exhibit a steeper decay in stimulus uncertainty, characterized by a larger exponent $\lambda$ (Fig.~\ref{fig:fig4}). In the mouse visual cortex and hippocampus, non-significant neurons lead to conditional entropy that decays less than half as fast as significant cells [Fig.~\ref{fig:fig4}(a)-(c)].

To understand how the coding properties of individual neurons impact the saturation of mutual information, we return to the realm of synthetic neurons. For generality, assume that the firing probability for each neuron-stimulus pair (denoted $\rho$) is drawn independently from an arbitrary distribution on the interval $[0,1]$. In the Supplemental Material, we show that the decay rate of the conditional entropy $H[s|\bm{x}]$ is given by
\begin{equation}
    \lambda
    = -\log\!\left(\langle\sqrt{\rho}\rangle^{2} + \langle\sqrt{1-\rho}\rangle^{2}\right).
\end{equation}
When neurons do not encode stimuli, the firing probabilities $\rho$ are constant, and $\lambda = 0$. By contrast, as $\rho$ becomes more variable, neural responses become more tightly tied to specific stimuli. In this case, the decay rate increases, and the mutual information saturates more quickly.

We can still compute the decay rate even when neurons have correlations in firing probabilities between different stimuli (Supplemental Material). The resulting decay is always slower than for independent firing probabilities. These results establish, both in experimental recordings and theory, that mutual information saturates exponentially with population size for conditionally independent neurons. \\

\emph{Discussion}---Quantifying how neural populations encode stimuli is central to understanding sensory and cognitive computation~\cite{Georgopoulos1988, Averbeck2006, Benucci2009}.
However, estimating population-level mutual information is notoriously difficult due to the exponential explosion of possible responses with population size.
Here, we use a Monte Carlo framework to estimate high-dimensional mutual information between population responses and discrete stimulus variables under conditional independence. 
We model the response distribution $P(\bm{x}|s)$ as the product of single-neuron firing probabilities, which can be reliably estimated from experimental data. We then sample stimulus-response pairs to estimate mutual information. While Monte Carlo methods have previously been used to study synthetic neurons~\cite{Yarrow2012, Huang2018}, here we directly infer conditionally independent models from large-scale recordings of neural activity.

We verify our mutual information estimates in a model of synthetic neurons, which we solve analytically. For empirical recordings of large populations, we apply finite-data corrections to account for sampling bias, demonstrating that high-dimensional mutual information can be reliably estimated with only a modest number of experimental samples. Across mouse visual cortex, mouse hippocampus, and salamander retina, we find that mutual information saturates exponentially with population size.

These results immediately open the door for future research in several directions.
First, our Monte Carlo estimates can be generalized to any model of population responses $P(\bm{x} | s)$.
As a natural starting point, here we focus on the conditionally independent model in Eq.~(\ref{eq:PCI}). However, neural populations often contain shared variability, or noise correlations, that are not captured by conditional independence~\cite{Hellwig2000,Holmgren2003,Brown2009,Stringer2019}.
Therefore, an important next step is to investigate richer models $P(\bm{x}| s)$, perhaps including recent maximum entropy models that scale to large populations~\cite{Carcamo2025, Lynn2025, lynn2025exact}.
Second, the framework is not tied to a specific brain region, species, recording modality, stimulus variable, or response state space.
It can be applied to any population activity for which one can define an external variable of interest, 
including sensory stimuli, spatial position, behavioral state, decision variables, or internal latent states.
More broadly, the same strategy could be used to estimate high-dimensional mutual information in other complex biological systems composed of many interacting units, 
where direct estimation of the full joint distribution is impossible.
Thus, this work provides a path toward extending information-theoretic tools to the scale of modern neural and biological experiments.

\section{Acknowledgments}
The authors thank members of the Lynn lab for suggestions. This work was supported in part by Gruber Science Fellowship (Z.L.), Mossman and NSERC Postdoctoral Fellowships (M.P.L.), and by support from the National Institutes of Health (NIH/NIGMS R35GM160188) and the Department of Physics, Quantitative Biology Institute, and Wu Tsai Institute at Yale University (C.W.L).

\bibliography{references.bib}

\section{Appendix}
\emph{\label{app:method_preprocess}Neural data}---
All data were recorded in previous experiments. For mouse visual cortex, activity was recorded using two-photon calcium imaging ~\cite{Stringer2019}. 
We focus on two recordings:  8,991 neurons responding to orientation gratings and 11,523 neurons responding to natural images. For both recordings, neural responses were binarized relative to spontaneous activity by thresholding each neuron's stimulus response at \(\mu_i^{\mathrm{spont}}+2\sigma_i^{\mathrm{spont}}\), 
where \(\mu_i^{\mathrm{spont}}\) and \(\sigma_i^{\mathrm{spont}}\) denote the mean and standard deviation of spontaneous fluorescence for neuron \(i\).

For mouse hippocampus, activity was recorded for a population of 1,485 neurons using two-photon calcium imaging as the animal ran along a circular track~\cite{Gauthier2018}. Mice preparation, neuron imaging, cell identification, activity normalization, and binarization are previously described~\cite{Gauthier2018}. 
We discretized the position of the mouse into 32 bins of equal probability, which served as the stimulus states in our analysis. For salamander retina, activity was recorded for a population of 53 neurons using a multi-electrode array during presentation of a stochastically moving horizontal bar~\cite{Palmer2015}. Recording and spike-sorting procedures are described previously~\cite{Palmer2015}. 
We treated the instantaneous bar position as the stimulus, which we discretized into bins of equal probability.

For the visual cortex and hippocampus datasets, we divided neurons into two groups based on whether the responded significantly to external stimuli or not~\cite{Shi2026}. 
For each neuron, firing probabilities $P(x_i = 1|s)$ were compared with null probabilities obtained by circularly shifting activity traces in time. Firing probabilities above the 95${}^\text{th}$ percentile of nulls were considered significant. We then labeled roughly 20\% of neurons significant based on the number of stimuli to which the responded significantly. For the orientation gratings, we selected the 2,651 neurons that responded significantly to at least 7 stimuli. For the natural images, we selected the 2,982 neurons that responded significantly to at least 6 stimuli. For the hippocampus, we selected the 321 neurons that responded significantly to at least 4 binned locations.

\section{Supplemental Material}

\setcounter{figure}{0}
\renewcommand{\thefigure}{S\arabic{figure}}
\setcounter{table}{0}
\renewcommand{\thetable}{S\arabic{table}}
\setcounter{equation}{0}
\renewcommand{\theequation}{S\arabic{equation}}

\subsection{Monte-Carlo Estimation Bias}
Here we derive the scaling of the bias of the mutual information estimator due to finite data used to estimate the parameters of the conditional firing probabilities. 

Let $T_f$ be the number of data points used to estimate the conditional firing probability distributions used to compute the mutual information. To estimate the bias, we treat the estimated mutual information $\hat{I}_{T_f}=f(\hat{\theta}_{T_f})$ 
as a function of the parameters $\hat{\theta}_{T_f}$, 
and expand the expression around the true parameters $\theta^*$:
\begin{equation}
    \begin{aligned}
        f(\hat{\theta}_{T_f}) &= f(\theta^\ast) + \nabla f(\theta^\ast)^{\top}(\hat{\theta}_{T_f} - \theta^\ast)\\
            &\text{  }+ \frac{1}{2}(\hat{\theta}_{T_f} - \theta^\ast)^{\top}\mathbf{H}_f(\theta^\ast)(\hat{\theta}_{T_f} - \theta^\ast)
            + \cdots 
    \end{aligned}
\end{equation}
where $\mathbf{H}_f$ is the Hessian matrix of second derivatives. 
The expectation value of this bias is:
\begin{equation}
    \begin{aligned}
        E[f(\hat{\theta}_{T_f})] &=  
        f(\theta^\ast)
        + \nabla f(\theta^\ast)^{\top} E[\hat{\theta}_{T_f} - \theta^\ast] \\
        &\text{ }+ \frac{1}{2}E[tr(\mathbf{H}_f(\theta^\ast)(\hat{\theta}_{T_f} - \theta^\ast)(\hat{\theta}_{T_f} - \theta^\ast)^\top)] \\ 
        &= I(\theta^\ast)
        +0
        +\frac{1}{2}tr(\mathbf{H}_f E[(\hat{\theta}_{T_f} - \theta^\ast)(\hat{\theta}_{T_f} - \theta^\ast)^\top]) \\ 
        &= I(\theta^\ast)+\frac{1}{2T_f}tr(\mathbf{H}_f(\theta^\ast) \mathcal{I}^{-1}(\theta^*)).
    \end{aligned}
\end{equation}
Assuming that the Hessian evaluated at $\theta^\ast$ is approximately equal, or at least proportional, to the Fisher information matrix $\mathcal{I}(\cdot)$,
this gives an estimation for the bias scaling:
\begin{equation}
    \begin{aligned}
        E[f(\hat{\theta}_{T_f})-I^\ast] &\propto \frac{d}{2T_f} = \frac{K(N+1)}{2T_f}.
    \end{aligned}
\end{equation}
Here $d=K(N+1)$ is the number of parameters.

\subsection{Random Encoding Model}
Here we provide full details of the analytically-tractable model we construct to study conditionally-independent random neural encoding. 

Consider a stimulus $s$ taking $K$ discrete values with uniform probability $1/K$. 
The stimulus is encoded in the activity of $N$ conditionally independent neurons, 
where neuron $i$ fires on stimulus $j$ with probability $\rho_{ij} \in [0,1]$. 
The firing probabilities $\rho_{ij}$ are i.i.d. draws from a symmetric Beta distribution 
$p(\rho) \sim \text{Beta}(\beta, \beta)$, fully characterizing the system via a 
$N \times K$ random matrix $\boldsymbol{\rho}$.

The stimulus entropy is $H[s] = \ln K$, and the conditional probability of a neural 
response $\bm{x} \in \{0,1\}^N$ given stimulus $s_j$ is:
\begin{equation}
    p(\bm{x}|s_j) = \prod_{i=1}^{N} \rho_{ij}^{x_i}(1-\rho_{ij})^{1-x_i}.
\end{equation}

We compute the mutual information $\langle I[\bm{x};s] \rangle_{p(\boldsymbol{\rho})}$, 
averaged (quenched) over the random encoding matrix $\boldsymbol{\rho}$.
The quenched average of the conditional entropy simplifies as:
\begin{equation}
    \langle H[\bm{x}|s] \rangle_{p(\boldsymbol{\rho})} 
    = N \langle h_b(\rho) \rangle_{p(\rho)},
\end{equation}
where $h_b(y) = -y\ln y - (1-y)\ln(1-y)$ is the binary entropy function, 
using conditional independence of neurons and identical distribution of $\rho_{ij}$.

The marginal response probability is:
\begin{equation}
    p(\bm{x}) = \frac{1}{K}\sum_{j=1}^{K} p(\bm{x}|s_j).
\end{equation}
Expanding the quenched average of the marginal entropy 
$\langle H[\bm{x}] \rangle_{p(\boldsymbol{\rho})}$, we separate it into two terms:
\begin{align}
    \langle H[\bm{x}] \rangle_{p(\boldsymbol{\rho})} 
    &= \frac{\ln K}{K} \sum_{\bm{x}} 
       \left\langle \sum_{j=1}^{K} p(\bm{x}|s_j) \right\rangle_{p(\boldsymbol{\rho})}
    \tag{Term 1} \\
    &- \frac{1}{K} \sum_{\bm{x}} 
       \left\langle \left(\sum_{j=1}^{K} p(\bm{x}|s_j)\right) 
       \ln\left(\sum_{j=1}^{K} p(\bm{x}|s_j)\right) \right\rangle_{p(\boldsymbol{\rho})}.
    \tag{Term 2}
\end{align}

Term 1 evaluates to $\ln K$ by normalization of the joint distribution $p(\bm{x}, s_j)$. 
For Term 2, factoring out the $j=1$ stimulus and defining the log-likelihood ratio 
$I_2 \equiv \left\langle \ln\!\left(1 + \sum_{\ell \neq j} 
\frac{p(\bm{x}|s_\ell)}{p(\bm{x}|s_j)}\right) \right\rangle$, 
we obtain:
\begin{equation}
    \langle H[\bm{x}] \rangle_{p(\boldsymbol{\rho})} 
    = \ln K + \langle H[\bm{x}|s] \rangle_{p(\boldsymbol{\rho})} - I_2,
\end{equation}
so that the mutual information is:
\begin{equation}
    \langle I[\bm{x};s] \rangle_{p(\boldsymbol{\rho})} = \ln K - I_2,
\end{equation}
with $\langle H[s|\bm{x}] \rangle_{p(\boldsymbol{\rho})} = I_2$.

To compute $I_2$ numerically, we apply the Frullani integral identity 
$\ln(A/B) = \int_0^\infty (e^{-Bt} - e^{-At})/t \, \mathrm{d}t$, 
setting $A = \sum_{\ell=1}^{K} p(\bm{x}|s_\ell)$ and $B = p(\bm{x}|s_j)$. 
Using independence of $p(\bm{x}|s_j)$ and $p(\bm{x}|s_\ell)$ for $\ell \neq j$, 
and defining:
\begin{equation}
    \phi(\bm{x}, t) \equiv \left\langle e^{-p(\bm{x}|s)t} \right\rangle_{p(\rho)},
\end{equation}
we obtain:
\begin{equation}
    I_2 = \sum_{\bm{x} \in \{0,1\}^N} \int_0^\infty \frac{\mathrm{d}t}{t} \,
    \phi'(\bm{x},t)\left[\phi(\bm{x},t)^{K-1} - 1\right].
\end{equation}

Since $p(\rho) \sim \text{Beta}(\beta,\beta)$ is symmetric about $\rho = 1/2$, 
we have $\rho \overset{d}{=} 1 - \rho$, so $\phi(\bm{x}, t)$ is independent 
of the specific pattern $\bm{x}$ and depends only on $N$. The expressions then simplify to:
\begin{align}
    I_2 &= 2^N \int_0^\infty \frac{\mathrm{d}t}{t} \,
    \phi'_N(t)\left[\phi_N(t)^{K-1} - 1\right], \\
    \phi_N(t) &= \sum_{k=0}^{\infty} \frac{(-t)^k}{k!} 
    \left\langle \rho^k \right\rangle^N,
\end{align}
where the moments $\langle \rho^k \rangle$ are evaluated under $\text{Beta}(\beta,\beta)$:
\begin{equation}
    \langle \rho^k \rangle = \frac{\prod_{r=0}^{k-1}(\beta + r)}{\prod_{r=0}^{k-1}(2\beta + r)}.
\end{equation}
This closed form enables exact numerical computation of the ground truth mutual information 
$\langle I[\bm{x};s] \rangle_{p(\boldsymbol{\rho})}$ for any $\beta$ and $N$.

\subsection{Scaling with the Number of Neurons}
We will now show that the average conditional entropy \(\langle H[s|\bm{x}]\rangle\) decays exponentially with \(N\) in the large-\(N\) limit for arbitrary \(p(\rho)\). 
To do so, we return to the simplest expression for \(\langle H[s|\bm{x}]\rangle\),
\begin{align}
\langle H[s|\bm{x}]\rangle
&= I_2 \\
&= \left\langle
\ln\!\left(
1+\sum_{\ell\neq j}\frac{p(\bm{x}|s_\ell)}{p(\bm{x}|s_j)}
\right)
\right\rangle_{p(\bm{x}|s_j),\,p(s_j),\,p(\rho)} .
\end{align}

Without loss of generality, we set \(j=1\) and define
\begin{equation}
\Lambda_\ell
=
\ln \frac{p(\bm{x}|s_\ell)}{p(\bm{x}|s_1)} ,
\end{equation}
so that
\begin{equation}
I_2
=
\left\langle
\ln\!\left(
1+\sum_{\ell=2}^{K} e^{\Lambda_\ell}
\right)
\right\rangle .
\end{equation}

We then define the quantity
\begin{equation}
J_k
\equiv
\left\langle
\ln\!\left(1+e^{\Lambda_k}\right)
\right\rangle ,
\end{equation}
under the assumption that $\rho_{ij}$ are drawn i.i.d.\ from $p(\rho)$ across stimuli $j$, 
the distribution of $\Lambda_k$ is the same for every $k \neq 1$.
Therefore $J_k$ is independent of $k$, and we may denote it simply by $J_N$.
This satisfies \(J_N \le I_2\), since
\begin{equation}
\ln(1+e^{\Lambda_k})
\le
\ln\!\left(1+\sum_{\ell=2}^{K}e^{\Lambda_\ell}\right).
\end{equation}
We also note that
\begin{equation}
1+\sum_{\ell=2}^{K}e^{\Lambda_\ell}
\le
\prod_{\ell=2}^{K}(1+e^{\Lambda_\ell}),
\end{equation}

In the i.i.d. firing case where the firing probabilities of each neuron to each stimulus is drawn i.i.d. from the distribution $p(\rho)$,
we have \(I_2 \le (K-1)J_N\).
Thus we have a pair of inequalities sandwiching \(I_2\),
\begin{equation}
J_N \le I_2 \le (K-1)J_N .
\end{equation}
By the Sandwich theorem, it follows that
\begin{equation}\label{eq:sandwichtheoremresult}
\lim_{N\to\infty}\frac{1}{N}\ln I_2
=
\lim_{N\to\infty}\frac{1}{N}\ln J_N ,
\end{equation}
so long as \(K\) remains finite as \(N\to\infty\). Thus we need only determine the large-\(N\) scaling of \(J_N\).

Recall the definition of \(\Lambda_k\),
\begin{align}\label{eq:lambdakdef}
\Lambda_{k}
&=
\ln \prod_{i=1}^{N}
\frac{
\rho_{ik}^{x_i}(1-\rho_{ik})^{1-x_i}
}{
\rho_{i1}^{x_i}(1-\rho_{i1})^{1-x_i}
}
\nonumber\\
&=
\sum_{i=1}^{N}
\underbrace{\left[
x_i \ln\frac{\rho_{ik}}{\rho_{i1}}
+
(1-x_i)\ln\frac{1-\rho_{ik}}{1-\rho_{i1}}
\right]}_{\gamma_i} \\
&=
\sum_{i=1}^{N}\gamma_i ,
\end{align}
where we have defined the random variable \(\gamma_i\).

We will now derive upper and lower bounds on \(J_N\) using the statistics of \(\gamma_i\).
First note that for any \(s\in(0,1)\), there exists a finite constant \(C_s\) such that
\[
\ln(1+e^u)\le C_s e^{su}
\qquad \text{for all real } u.
\]
This implies that
\begin{align}\label{eq:JNUB}
J_N
&\le
C_s \left\langle e^{s\sum_{i=1}^{N}\gamma_i}\right\rangle
=
C_s \left\langle e^{s\gamma_i}\right\rangle^N ,
\end{align}
where in the last step we used the independence of the \(\gamma_i\) for different neuron \(i\). 
Taking the limiting scaling behavior, and noting that \(C_s\) is independent of \(N\), we obtain
\begin{equation}
\lim_{N\to\infty}\frac{1}{N}\ln J_N
\le
\ln \langle e^{s\gamma_i}\rangle
\equiv
\ln M(s).
\end{equation}

We can evaluate \(M(s)\) by noting that, given stimulus $s_1$, 
we have \( x_i=1\) with probability \(\rho_{i1}\) and \( x_i=0\) with probability \(1-\rho_{i1}\). 
Thus
\begin{align}\label{eq:Msdef}
M(s)
&=
\left\langle
\rho_{i1}\left(\frac{\rho_{ik}}{\rho_{i1}}\right)^s
+
(1-\rho_{i1})
\left(
\frac{1-\rho_{ik}}{1-\rho_{i1}}
\right)^s
\right\rangle_{p(\rho)}
\nonumber\\
&=
\langle \rho^{1-s}\rangle \langle \rho^s\rangle
+
\langle (1-\rho)^{1-s}\rangle \langle (1-\rho)^s\rangle .
\end{align}

Since the upper bound holds for all \(s\in(0,1)\), it also holds for the minimizing \(s\), so that
\begin{equation}
\lim_{N\to\infty}\frac{1}{N}\ln J_N
\le
\min_{s\in(0,1)} \ln M(s).
\end{equation}

We now derive a complementary lower bound. First note that
\begin{equation}
\ln(1+e^u)\ge \ln 2 \, H(u),
\end{equation}
where \(H(u)\) is the Heaviside step function. Therefore,
\begin{equation}
J_N
\ge
\ln 2 \, P(\Lambda_k\ge 0).
\end{equation}

To evaluate this probability we use Cram\'er's theorem. Let
\begin{equation}
\psi(s)=\ln M(s),
\end{equation}
which is a log-moment-generating function and therefore finite for all \(s\in\mathbb{R}\). Cram\'er's theorem states that
\begin{equation}
\lim_{N\to\infty}\frac{1}{N}
\ln
P\!\left(\sum_{i=1}^{N}\gamma_i \ge Nx\right)
=
-\psi^*(x)
=
-\sup_{s\in\mathbb{R}}\left[sx-\psi(s)\right].
\end{equation}

Noting that
\[
\psi^*(0)=-\inf_{s\in\mathbb{R}}\ln M(s),
\]
we have
\begin{equation}\label{eq:JNLB}
\lim_{N\to\infty}\frac{1}{N}\ln J_N
\ge
\inf_{s\in\mathbb{R}}\ln M(s).
\end{equation}

Combining the lower and upper bounds on \(J_N\) with the Sandwich theorem result, Eq.~\eqref{eq:sandwichtheoremresult}, we obtain
\begin{equation}
\inf_{s\in\mathbb{R}}\ln M(s)
\le
\lim_{N\to\infty}\frac{1}{N}\ln I_2
\le
\min_{s\in(0,1)}\ln M(s).
\end{equation}

We now show that the global minimum of \(M(s)\) occurs at \(s=\tfrac12\), so that the lower and upper bounds coincide. Our problem is therefore to find
\begin{equation}
\min_{s\in\mathbb{R}}
\ln\!\left(
\langle \rho^{1-s}\rangle \langle \rho^s\rangle
+
\langle (1-\rho)^{1-s}\rangle \langle (1-\rho)^s\rangle
\right).
\end{equation}

Consider the Cauchy--Schwarz inequality,
\[
\langle XY\rangle^2 \le \langle X^2\rangle \langle Y^2\rangle .
\]
Let \(X=\rho^{s/2}\) and \(Y=\rho^{(1-s)/2}\). Then
\begin{equation}
\langle \rho^{1/2}\rangle^2
\le
\langle \rho^{1-s}\rangle \langle \rho^s\rangle .
\end{equation}
By symmetry, similarly,
\begin{equation}
\langle (1-\rho)^{1/2}\rangle^2
\le
\langle (1-\rho)^{1-s}\rangle \langle (1-\rho)^s\rangle .
\end{equation}

It follows that $M(1/2)\le M(s)$, for all $s\in\mathbb{R},$
so the global minimum of \(M(s)\) lies at \(s=\tfrac12\). Hence the upper and lower bounds coincide, and we conclude that
\begin{align}
\lambda & \equiv -\lim_{N\to\infty}\frac{1}{N}\ln \langle H[s|\bm{x}]\rangle\\
&=
-\lim_{N\to\infty}\frac{1}{N}\ln I_2 \\
&=
-\ln\!\left(
\langle \sqrt{\rho}\rangle^2
+
\langle \sqrt{1-\rho}\rangle^2
\right).\label{eq:scalingresult}
\end{align}

\subsection{Correlated Firing Probabilities}
Now suppose the firing probabilities of a given neuron across stimuli are correlated. Denoting by $\rho_{ij}$ the firing probability of neuron $i$ on stimulus $j$, the vectors
$(\rho_{i1},\ldots,\rho_{iK})$ are drawn independently and
identically across neurons, with the same marginal distribution
$p(\rho)$. We assume each pair $(\rho_{ij},\rho_{ik})$
has a joint distribution invariant under exchanging its
entries, but this distribution may differ between pairs. The derivation largely follows the previous one for uncorrelated firing probabilities; we sketch the key steps below.

For a given pair of stimuli $j$ and $k$, define
\begin{equation}
    \Lambda_{jk}
    = \ln\frac{p(\bm{x}| s_k)}
               {p(\bm{x}| s_j)},
    \qquad
    J_{jk,N}
    = \left\langle\ln(1+e^{\Lambda_{jk}})\right\rangle_{p(\textbf{x}|s_j)p(\rho)}.
\end{equation}

As in Eqs.~\eqref{eq:lambdakdef}--\eqref{eq:Msdef}, $\Lambda_{jk}$ is a sum of
independent single-neuron contributions $\gamma_i^{jk}$,
with moment-generating function
\begin{equation}
    \begin{aligned}
        M_{jk}(s)
        &\equiv \left\langle e^{s\gamma_i^{jk}}\right\rangle_{p(\textbf{x}|s_j)p(\rho)}\\
        &= \Bigl\langle
            \rho_{ij}^{\,1-s}\rho_{ik}^{\,s} +(1-\rho_{ij})^{\,1-s}(1-\rho_{ik})^{\,s}
        \Bigr\rangle.
    \end{aligned}
\end{equation}

Pairwise symmetry gives $M_{jk}(s)=M_{jk}(1-s)$.
Applying Cauchy--Schwarz directly to the exponential
moments, as before, therefore yields
\begin{equation}
    M_{jk}(1/2)^2
    \leq M_{jk}(s)M_{jk}(1-s)
    = M_{jk}(s)^2.
\end{equation}
Thus the global minimum occurs at $s=1/2$, and the
upper and lower bounds of Eqs.~\eqref{eq:JNUB}--\eqref{eq:JNLB} coincide
for each pair, giving
$\lim_{N\to\infty}N^{-1}\ln J_{jk,N}
=\ln M_{jk}(1/2)$.

Finally, applying the Sandwich theorem as in
Eq.~\eqref{eq:sandwichtheoremresult} gives
\begin{equation}
    \frac{1}{K}\max_{j\neq k}J_{jk,N}
    \leq I_2
    \leq (K-1)\max_{j\neq k}J_{jk,N}.
\end{equation}
For fixed $K\geq2$, the slowest-decaying pairwise exponential
rate therefore determines the scaling:
\begin{equation}
    \begin{aligned}
        \lambda_{\mathrm{corr}}
        &\equiv
        -\lim_{N\to\infty}\frac{1}{N}
        \ln\left\langle H[s|\bm{x}]\right\rangle\\
        &=
        -\ln\!\Bigl[
            \max_{j<k}
            \Bigl\langle
                \sqrt{\rho_{ij}\rho_{ik}}+\sqrt{(1-\rho_{ij})(1-\rho_{ik})}
            \Bigr\rangle
        \Bigr].
    \end{aligned}
    \label{eq:corr-scaling-result}
\end{equation}
When firing probabilities are independent across stimuli, the joint moments factorize and are identical for every pair, recovering Eq.~\eqref{eq:scalingresult}.

\begin{figure*}
\includegraphics{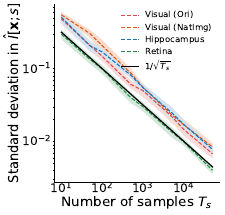}
\caption{\label{fig:supp1} Convergence of the Monte Carlo mutual information estimator with the number of samples \(T_s\). Results are shown for neuron-group size \(N=50\) in all four datasets.}
\end{figure*}

\begin{figure*}
\includegraphics{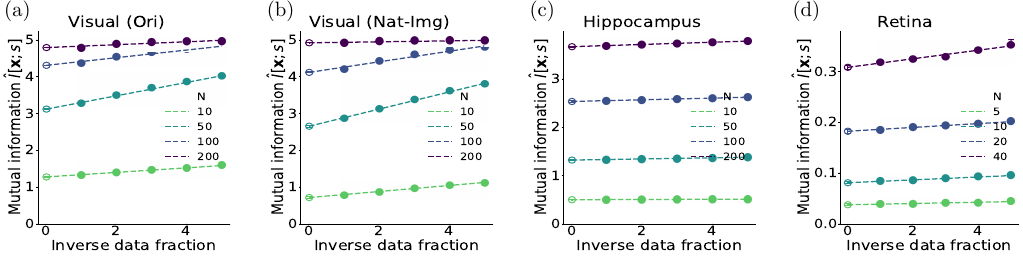}
\caption{\label{fig:supp2} Example finite data correction across empirical datasets.
\textbf{a}. Primary visual cortex data under orientation stimulation.
\textbf{b}. Primary visual cortex data under natural-image stimulation. 
\textbf{c}. Hippocampus data. 
\textbf{d}. Retinal data.
}
\end{figure*}

\begin{figure*}
\includegraphics{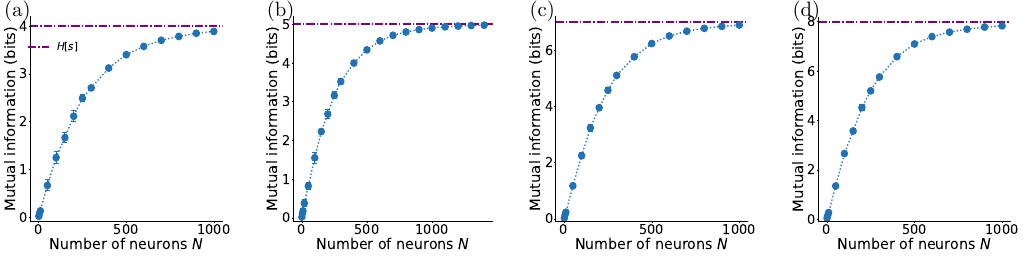}
\caption{\label{fig:supp3} Discretization on continuous stimulus do not reflect qualitative difference in hippocampus mutual information estimation. 
\textbf{a}. $K=16$.
\textbf{b}. $K=32$. 
\textbf{c}. $K=128$. 
\textbf{d}. $K=256$.
}
\end{figure*}

\end{document}